\documentclass[twocolumn,letterpaper,aps,prc,longbibliography,superscriptaddress,nofootinbib,floatfix]{revtex4-1}
\usepackage{graphicx}	
\usepackage{xspace}	

\newcommand{\pt}{\mbox{$p_T$}\xspace}
\newcommand{\piz}{\mbox{$\pi^0$}\xspace}
\newcommand{\gevc}{\mbox{GeV/$c$}\xspace}

\newcommand{\rab}{\mbox{$R_{AB}$}\xspace}

\newcommand{\tab}{\mbox{$T_{AB}$}\xspace}
\newcommand{\Npart}{\mbox{$N_{\rm part}$}\xspace}
\newcommand{\Ncoll}{\mbox{$N_{\rm coll}$}\xspace}

\newcommand{\pp}{\mbox{$p$$+$$p$}\xspace}

\newcommand{\auau}{\mbox{Au$+$Au}\xspace}

\newcommand{\cucu}{\mbox{Cu$+$Cu}\xspace}
\newcommand{\cuau}{\mbox{Cu$+$Au}\xspace}

\newcommand{\aaa}{\mbox{A$+$A}\xspace}

\begin{document}

\title{Production of $\pi^0$ and $\eta$ mesons in Cu$+$Au collisions at 
$\sqrt{s_{_{NN}}}$=200 GeV}

\newcommand{\abilene}{Abilene Christian University, Abilene, Texas 79699, USA}
\newcommand{\augie}{Department of Physics, Augustana University, Sioux Falls, South Dakota 57197, USA}
\newcommand{\banaras}{Department of Physics, Banaras Hindu University, Varanasi 221005, India}
\newcommand{\barc}{Bhabha Atomic Research Centre, Bombay 400 085, India}
\newcommand{\baruch}{Baruch College, City University of New York, New York, New York, 10010 USA}
\newcommand{\bnlcoll}{Collider-Accelerator Department, Brookhaven National Laboratory, Upton, New York 11973-5000, USA}
\newcommand{\bnlphys}{Physics Department, Brookhaven National Laboratory, Upton, New York 11973-5000, USA}
\newcommand{\caucr}{University of California-Riverside, Riverside, California 92521, USA}
\newcommand{\charlesczech}{Charles University, Ovocn\'{y} trh 5, Praha 1, 116 36, Prague, Czech Republic}
\newcommand{\chonbuk}{Chonbuk National University, Jeonju, 561-756, Korea}
\newcommand{\ciae}{Science and Technology on Nuclear Data Laboratory, China Institute of Atomic Energy, Beijing 102413, People's Republic of China}
\newcommand{\cns}{Center for Nuclear Study, Graduate School of Science, University of Tokyo, 7-3-1 Hongo, Bunkyo, Tokyo 113-0033, Japan}
\newcommand{\colorado}{University of Colorado, Boulder, Colorado 80309, USA}
\newcommand{\columbia}{Columbia University, New York, New York 10027 and Nevis Laboratories, Irvington, New York 10533, USA}
\newcommand{\czechtech}{Czech Technical University, Zikova 4, 166 36 Prague 6, Czech Republic}
\newcommand{\debrecen}{Debrecen University, H-4010 Debrecen, Egyetem t{\'e}r 1, Hungary}
\newcommand{\elte}{ELTE, E{\"o}tv{\"o}s Lor{\'a}nd University, H-1117 Budapest, P{\'a}zm{\'a}ny P.~s.~1/A, Hungary}
\newcommand{\eszterhazy}{Eszterh\'azy K\'aroly University, K\'aroly R\'obert Campus, H-3200 Gy\"ongy\"os, M\'atrai \'ut 36, Hungary}
\newcommand{\ewha}{Ewha Womans University, Seoul 120-750, Korea}
\newcommand{\fsu}{Florida State University, Tallahassee, Florida 32306, USA}
\newcommand{\gsu}{Georgia State University, Atlanta, Georgia 30303, USA}
\newcommand{\hanyang}{Hanyang University, Seoul 133-792, Korea}
\newcommand{\hiroshima}{Hiroshima University, Kagamiyama, Higashi-Hiroshima 739-8526, Japan}
\newcommand{\howard}{Department of Physics and Astronomy, Howard University, Washington, DC 20059, USA}
\newcommand{\ihepprot}{IHEP Protvino, State Research Center of Russian Federation, Institute for High Energy Physics, Protvino, 142281, Russia}
\newcommand{\illuiuc}{University of Illinois at Urbana-Champaign, Urbana, Illinois 61801, USA}
\newcommand{\inrras}{Institute for Nuclear Research of the Russian Academy of Sciences, prospekt 60-letiya Oktyabrya 7a, Moscow 117312, Russia}
\newcommand{\instpasczech}{Institute of Physics, Academy of Sciences of the Czech Republic, Na Slovance 2, 182 21 Prague 8, Czech Republic}
\newcommand{\isu}{Iowa State University, Ames, Iowa 50011, USA}
\newcommand{\jaea}{Advanced Science Research Center, Japan Atomic Energy Agency, 2-4 Shirakata Shirane, Tokai-mura, Naka-gun, Ibaraki-ken 319-1195, Japan}
\newcommand{\jyvaskyla}{Helsinki Institute of Physics and University of Jyv{\"a}skyl{\"a}, P.O.Box 35, FI-40014 Jyv{\"a}skyl{\"a}, Finland}
\newcommand{\kek}{KEK, High Energy Accelerator Research Organization, Tsukuba, Ibaraki 305-0801, Japan}
\newcommand{\korea}{Korea University, Seoul, 136-701, Korea}
\newcommand{\kurchatov}{National Research Center ``Kurchatov Institute", Moscow, 123098 Russia}
\newcommand{\kyoto}{Kyoto University, Kyoto 606-8502, Japan}
\newcommand{\labllr}{Laboratoire Leprince-Ringuet, Ecole Polytechnique, CNRS-IN2P3, Route de Saclay, F-91128, Palaiseau, France}
\newcommand{\lahorelums}{Physics Department, Lahore University of Management Sciences, Lahore 54792, Pakistan}
\newcommand{\lawllnl}{Lawrence Livermore National Laboratory, Livermore, California 94550, USA}
\newcommand{\losalamos}{Los Alamos National Laboratory, Los Alamos, New Mexico 87545, USA}
\newcommand{\lund}{Department of Physics, Lund University, Box 118, SE-221 00 Lund, Sweden}
\newcommand{\lyon}{IPNL, CNRS/IN2P3, Univ Lyon, Université Lyon 1, F-69622, Villeurbanne, France}
\newcommand{\maryland}{University of Maryland, College Park, Maryland 20742, USA}
\newcommand{\mass}{Department of Physics, University of Massachusetts, Amherst, Massachusetts 01003-9337, USA}
\newcommand{\michigan}{Department of Physics, University of Michigan, Ann Arbor, Michigan 48109-1040, USA}
\newcommand{\muhlenberg}{Muhlenberg College, Allentown, Pennsylvania 18104-5586, USA}
\newcommand{\myongji}{Myongji University, Yongin, Kyonggido 449-728, Korea}
\newcommand{\nagasaki}{Nagasaki Institute of Applied Science, Nagasaki-shi, Nagasaki 851-0193, Japan}
\newcommand{\nara}{Nara Women's University, Kita-uoya Nishi-machi Nara 630-8506, Japan}
\newcommand{\natmephi}{National Research Nuclear University, MEPhI, Moscow Engineering Physics Institute, Moscow, 115409, Russia}
\newcommand{\newmex}{University of New Mexico, Albuquerque, New Mexico 87131, USA}
\newcommand{\nmsu}{New Mexico State University, Las Cruces, New Mexico 88003, USA}
\newcommand{\ohio}{Department of Physics and Astronomy, Ohio University, Athens, Ohio 45701, USA}
\newcommand{\ornl}{Oak Ridge National Laboratory, Oak Ridge, Tennessee 37831, USA}
\newcommand{\orsay}{IPN-Orsay, Univ.~Paris-Sud, CNRS/IN2P3, Universit\'e Paris-Saclay, BP1, F-91406, Orsay, France}
\newcommand{\peking}{Peking University, Beijing 100871, People's Republic of China}
\newcommand{\pnpi}{PNPI, Petersburg Nuclear Physics Institute, Gatchina, Leningrad region, 188300, Russia}
\newcommand{\riken}{RIKEN Nishina Center for Accelerator-Based Science, Wako, Saitama 351-0198, Japan}
\newcommand{\rikjrbrc}{RIKEN BNL Research Center, Brookhaven National Laboratory, Upton, New York 11973-5000, USA}
\newcommand{\rikkyo}{Physics Department, Rikkyo University, 3-34-1 Nishi-Ikebukuro, Toshima, Tokyo 171-8501, Japan}
\newcommand{\saispbstu}{Saint Petersburg State Polytechnic University, St.~Petersburg, 195251 Russia}
\newcommand{\seoulnat}{Department of Physics and Astronomy, Seoul National University, Seoul 151-742, Korea}
\newcommand{\stonybrkc}{Chemistry Department, Stony Brook University, SUNY, Stony Brook, New York 11794-3400, USA}
\newcommand{\stonycrkp}{Department of Physics and Astronomy, Stony Brook University, SUNY, Stony Brook, New York 11794-3800, USA}
\newcommand{\sungskku}{Sungkyunkwan University, Suwon, 440-746, Korea}
\newcommand{\tenn}{University of Tennessee, Knoxville, Tennessee 37996, USA}
\newcommand{\titech}{Department of Physics, Tokyo Institute of Technology, Oh-okayama, Meguro, Tokyo 152-8551, Japan}
\newcommand{\tsukuba}{Tomonaga Center for the History of the Universe, University of Tsukuba, Tsukuba, Ibaraki 305, Japan}
\newcommand{\vandy}{Vanderbilt University, Nashville, Tennessee 37235, USA}
\newcommand{\weizmann}{Weizmann Institute, Rehovot 76100, Israel}
\newcommand{\wigner}{Institute for Particle and Nuclear Physics, Wigner Research Centre for Physics, Hungarian Academy of Sciences (Wigner RCP, RMKI) H-1525 Budapest 114, POBox 49, Budapest, Hungary}
\newcommand{\yonsei}{Yonsei University, IPAP, Seoul 120-749, Korea}
\newcommand{\zagreb}{Department of Physics, Faculty of Science, University of Zagreb, Bijeni\v{c}ka c.~32 HR-10002 Zagreb, Croatia}
\affiliation{\abilene}
\affiliation{\augie}
\affiliation{\banaras}
\affiliation{\barc}
\affiliation{\baruch}
\affiliation{\bnlcoll}
\affiliation{\bnlphys}
\affiliation{\caucr}
\affiliation{\charlesczech}
\affiliation{\chonbuk}
\affiliation{\ciae}
\affiliation{\cns}
\affiliation{\colorado}
\affiliation{\columbia}
\affiliation{\czechtech}
\affiliation{\debrecen}
\affiliation{\elte}
\affiliation{\eszterhazy}
\affiliation{\ewha}
\affiliation{\fsu}
\affiliation{\gsu}
\affiliation{\hanyang}
\affiliation{\hiroshima}
\affiliation{\howard}
\affiliation{\ihepprot}
\affiliation{\illuiuc}
\affiliation{\inrras}
\affiliation{\instpasczech}
\affiliation{\isu}
\affiliation{\jaea}
\affiliation{\jyvaskyla}
\affiliation{\kek}
\affiliation{\korea}
\affiliation{\kurchatov}
\affiliation{\kyoto}
\affiliation{\labllr}
\affiliation{\lahorelums}
\affiliation{\lawllnl}
\affiliation{\losalamos}
\affiliation{\lund}
\affiliation{\lyon}
\affiliation{\maryland}
\affiliation{\mass}
\affiliation{\michigan}
\affiliation{\muhlenberg}
\affiliation{\myongji}
\affiliation{\nagasaki}
\affiliation{\nara}
\affiliation{\natmephi}
\affiliation{\newmex}
\affiliation{\nmsu}
\affiliation{\ohio}
\affiliation{\ornl}
\affiliation{\orsay}
\affiliation{\peking}
\affiliation{\pnpi}
\affiliation{\riken}
\affiliation{\rikjrbrc}
\affiliation{\rikkyo}
\affiliation{\saispbstu}
\affiliation{\seoulnat}
\affiliation{\stonybrkc}
\affiliation{\stonycrkp}
\affiliation{\sungskku}
\affiliation{\tenn}
\affiliation{\titech}
\affiliation{\tsukuba}
\affiliation{\vandy}
\affiliation{\weizmann}
\affiliation{\wigner}
\affiliation{\yonsei}
\affiliation{\zagreb}
\author{C.~Aidala} \affiliation{\losalamos} \affiliation{\michigan} 
\author{N.N.~Ajitanand} \altaffiliation{Deceased} \affiliation{\stonybrkc} 
\author{Y.~Akiba} \email[PHENIX Spokesperson: ]{akiba@rcf.rhic.bnl.gov} \affiliation{\riken} \affiliation{\rikjrbrc} 
\author{R.~Akimoto} \affiliation{\cns} 
\author{J.~Alexander} \affiliation{\stonybrkc} 
\author{M.~Alfred} \affiliation{\howard} 
\author{K.~Aoki} \affiliation{\kek} \affiliation{\riken} 
\author{N.~Apadula} \affiliation{\isu} \affiliation{\stonycrkp} 
\author{H.~Asano} \affiliation{\kyoto} \affiliation{\riken} 
\author{E.T.~Atomssa} \affiliation{\stonycrkp} 
% ...
\collaboration{PHENIX Collaboration} \noaffiliation

\date{\today}


\begin{abstract}

Production of $\pi^0$ and $\eta$ mesons has been measured at midrapidity 
in Cu$+$Au collisions at $\sqrt{s_{_{NN}}}$=200 GeV. Measurements were 
performed in $\pi^0(\eta)\rightarrow\gamma\gamma$ decay channel in the 
1(2)-20 GeV/$c$ transverse momentum range. A strong suppression is 
observed for $\pi^0$ and $\eta$ meson production at high transverse 
momentum in central Cu$+$Au collisions relative to the $p$$+$$p$ results 
scaled by the number of nucleon-nucleon collisions. In central 
collisions the suppression is similar to Au$+$Au with comparable nuclear 
overlap. The $\eta/\pi^0$ ratio measured as a function of transverse 
momentum is consistent with $m_T$-scaling parameterization down to 
$p_T=2$ GeV/$c$, its asymptotic value is constant and consistent with 
Au$+$Au and $p$$+$$p$ and does not show any significant dependence on 
collision centrality. Similar results were obtained in hadron-hadron, 
hadron-nucleus, and nucleus-nucleus collisions as well as in $e^+e^-$ 
collisions in a range of collision energies $\sqrt{s_{_{NN}}}=3$--1800 
GeV.  This suggests that the quark-gluon-plasma medium produced in 
Cu$+$Cu collisions either does not affect the jet fragmentation into 
light mesons or it affects the $\pi^0$ and $\eta$ the same way.

\end{abstract}

\maketitle

\section{Introduction}

Experiments at the Relativistic Heavy Ion Collider 
(RHIC)~\cite{Arsene:2004fa,Back:2004je,Adams:2005dq,Adcox:2004mh} and 
later at the Large Hadron 
Collider (LHC)~\cite{CMS:2012aa,Abelev:2012hxa,Aad:2012vca,Gunji:2016uqe} 
established the formation of quark-gluon plasma (QGP) in relativistic 
collisions of heavy ions (\aaa). One of the most important tools to 
investigate the properties of this new medium are identified hadrons at 
high transverse momenta (\pt$>$5\,\gevc), because they are leading 
fragments~\cite{jacob1978large} of jets from hard-scattered partons, 
which, before fragmentation, interacted with the 
QGP~\cite{Baier:2000mf}. The differential cross section of high-$p_T$ 
hadron production in elementary $p$+$p$ collisions can be derived using 
next-to-leading-order perturbative-quantum-chromodynamics 
formalism~\cite{Owens:1986mp,Adare:2007dg}:

\begin{eqnarray}\label{eqcs}
d\sigma_{pp\rightarrow hX}\approx\sum_{abcd}\int 
dx_a dx_b dz_c f_{a/p}(x_a)\otimes f_{b/p}(x_b) \\ 
\otimes \  d\sigma_{ab\rightarrow cd} \otimes D_{c\rightarrow h}(z_c), \nonumber
\end{eqnarray}

\noindent where $x_{a,b}$ is the initial momentum fraction carried by 
partons $a$ and $b$, $z_c$ is the final-state momentum fraction of the 
hadron $h$, $f_{a/p}$ and $f_{b/p}$ are the parton distribution 
functions (PDFs), $d\sigma_{ab\rightarrow cd}$ is the differential cross 
section of the initial partons hard scattering, $D_{c\rightarrow h}$ is 
the fragmentation function (FF) of the hard scattered parton to the 
final-state hadron.

There are two classes of nuclear effects, which modify the high-$p_T$ 
hadron production cross section in A+A collisions. The initial state (or 
cold nuclear matter) effects are related to the presence of a heavy 
nucleus in the collision and require the modification of the 
corresponding PDF in Eq. \ref{eqcs}. The correction factors for PDFs are 
usually obtained from the $p$/$d$$+$$A$ 
data~\cite{Helenius:2012wd,Eskola:2016oht}.

The final-state effects are related to the formation of a hot, dense 
medium, QGP. While the hard-scattered parton propagates through the 
medium, it loses a fraction of its energy 
(jet-quenching)~\cite{Bjorken:1982tu,Baier:2000mf,Wang:1994fx} by gluon 
emission or elastic scatterings with the medium constituents.

Parton energy loss in the QGP is quantified with the jet transport 
coefficient $\hat{q}$, defined as the squared momentum exchange between 
the hard parton and the medium per unit path length~\cite{Baier:2000mf}. 
The relation of $\hat{q}$ to other medium parameters, such as temperature, 
shear viscosity and entropy, is indicative of the character of the 
coupling to the medium~\cite{majumder2007small}.

Several phenomenological 
models~\cite{qin2008gy,chen2011x,young2012martini,majumder2012suppression,xu2014azimuthal} 
were designed to estimate $\hat{q}$ based on $\rab$ measurements at RHIC 
and the LHC.  In the model calculations, the final state effects are usually 
accounted for by replacing $D_{c\rightarrow h}$ with the medium-modified 
FF $\widetilde{D}_{c\rightarrow h}$ in Eq.~\ref{eqcs}. Methods of 
$\widetilde{D}_{c\rightarrow h}$ estimation are specific for each parton 
energy model. Also several attempts were made to extract $\hat{q}$ using 
lattice QCD calculations~\cite{majumder2013calculating,Panero:2013pla}. 
The effects of the medium on particle production are usually quantified 
by the nuclear modification factor ($\rab$):
\begin{equation}
   R_{AB}^{\rm cent}(\pt) = \frac{1}{T_{AB}^{\rm cent}}\frac{dN_{AB}^{\rm cent}/d\pt}{d\sigma_{pp}/d\pt},
\end{equation}
where $dN_{AB}^{\rm cent}$ is the particle yield measured in A+B collisions 
for a given centrality class ($cent$), $d\sigma_{pp}$ is the 
cross section of the same particle measured in $\pp$ collisions at the 
same collision energy, $T_{AB}^{\rm cent}$ is the nuclear thickness function 
for the event class~\cite{Miller:2007ri}. The energy loss of 
hard-scattered partons causes a reduction of \rab from unity towards 
smaller values.

Measurement of $\pi^0$ meson production is particularly interesting, 
because $\pi^0$s are abundantly produced and their yields can be measured 
up to high $p_T$ with good particle identification, an excellent 
signal-to-background ratio ($S/B$), and relatively small uncertainties 
using the electromagnetic calorimeters (EMCal) of the PHENIX detector. The 
$\eta$ meson has four times heavier mass than $\pi^0$ and an about 50\% 
strangeness content.  Thus measurements of $\eta$ allow to study the 
dependence of jet quenching on the hadron mass and flavor content. 
Measurement of $\eta/\pi^0$ in $\aaa$ gives an opportunity to better 
understand whether fragmentation processes are affected by the presence 
of the colored medium.

\begin{figure}[htbp]
  \includegraphics[width=1.0\linewidth]{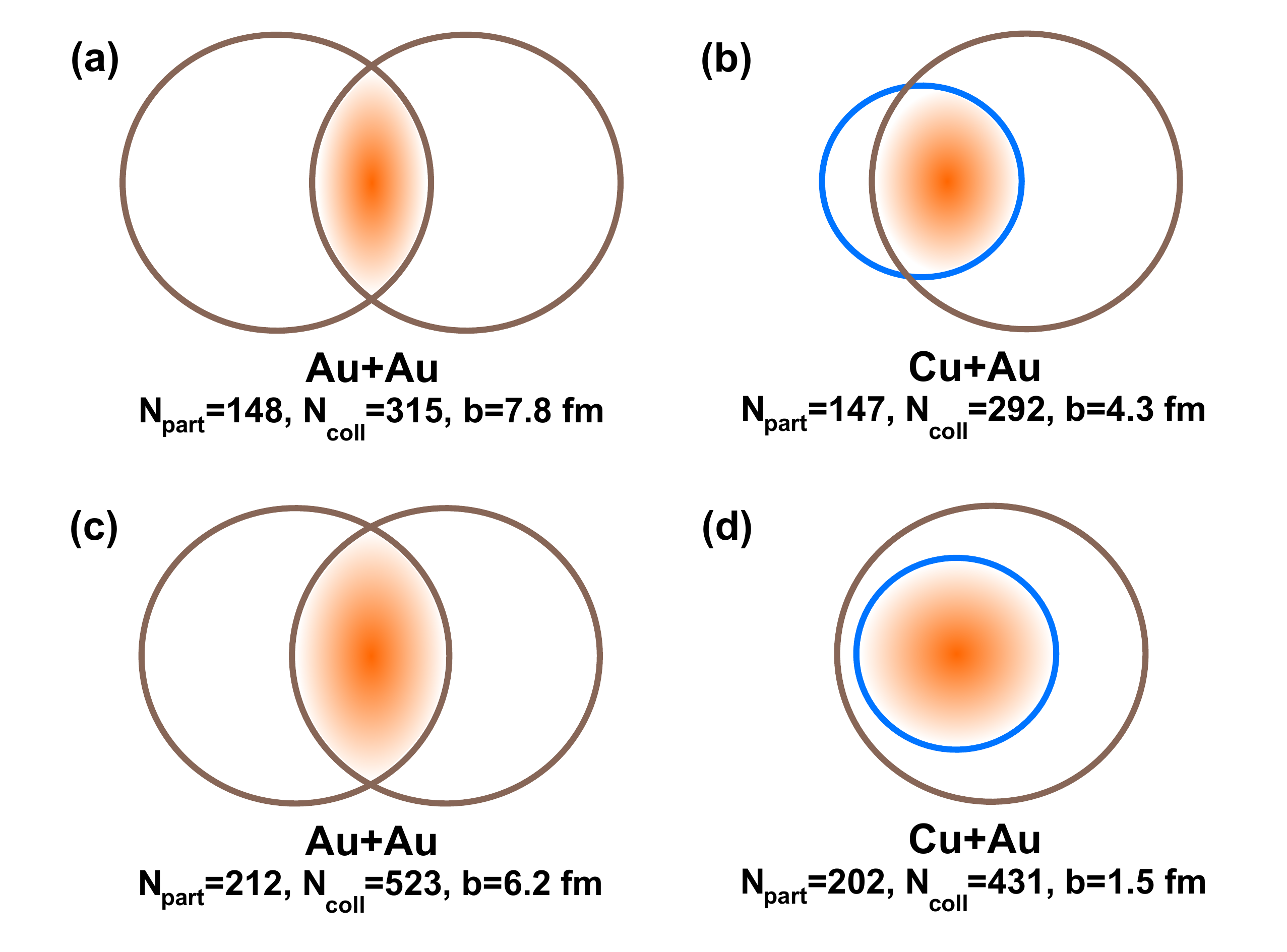}
\caption{Cartoon showing \auau and \cuau collisions with comparable 
\Npart. On panel (b) the overlap area is asymmetric, and part of the 
dilute surfaces of Au and Cu (corona) overlap.
}
\label{fig:cartoon}
\end{figure}

Previously published results on $\pi^0$ and $\eta$ production at PHENIX 
were obtained in symmetric heavy-ion systems such as $\auau$ and $\cucu$ 
~\cite{Adler:2006bv, Adare:2008cg, Adare:2008qa, Adare:2008ad, 
Adare:2010dc}. Contrarily to that, $\cuau$ collisions at 
$\sqrt{s_{_{NN}}}=200$ GeV is the first asymmetric system of heavy nuclei 
studied at RHIC. Such collisions provide a different collision geometry 
from the one realized in symmetric systems. In central collisions the Cu 
nucleus is fully submerged in the Au nucleus, which results in the 
reduction of nucleon-nucleon interactions in the ``corona"
region~\cite{Pantuev:2005jt} of the collision (see 
Fig.~\ref{fig:cartoon}). In semi-central $\cuau$ collisions an asymmetry 
of the nuclear overlap region is present along the axis connecting the 
centers of the interacting nuclei. These features make $\cuau$ collision 
system an important part of the systematic study of the final-state 
effects in heavy-ion collisions.

In this paper we present $\pi^0$ and $\eta$ meson $p_T$ spectra and 
nuclear modification factor measurements in $\cuau$ collisions at 
$\sqrt{s_{_{NN}}}=200$ GeV. Data were collected in RHIC Year-2012 run with 
the PHENIX detector.

\section{Data analysis}

\begin{table}[htb]
\caption{\label{tab0}
The average values of the nuclear thickness function $\left\langle 
T_{AB} \right\rangle$ and the numbers of nucleons participating in the 
nuclei interaction $\left\langle N_{\rm part} \right\rangle$ in different 
Cu$+$Au centrality intervals. }
\begin{ruledtabular} \begin{tabular}{ccc}
Centrality interval & $\left\langle T_{AB} \right\rangle$ (mb$^{-1}$) & $\left\langle N_{\rm part} \right\rangle$ \\
\hline
Minimum Bias & $2.54\pm 0.19$ & $61.1\pm 2.7$ \\
0\%--10\% & $8.8\pm 0.6$ & $177.2\pm 5.2$ \\
10\%--20\% & $6.0\pm 0.4$ & $132.4\pm 3.7$ \\
0\%--20\% & $7.5\pm 0.5$ & $154.8\pm 4.1$ \\
20\%--40\% & $3.1\pm 0.2$ & $80.4\pm 3.3$ \\
40\%--60\% & $1.00\pm 0.12$ & $34.9\pm 2.8$ \\
\end{tabular} \end{ruledtabular}
\end{table}

A detailed description of the PHENIX experimental set-up can be found 
elsewhere~\cite{Adcox:2003zm}. Beam-beam counters (BBC, 
$3.0<|\eta|<3.9$)~\cite{Allen:2003zt} located downstream in both beam 
directions (north and south), each consisting of 64 \v{C}erenkov-radiator 
counters, provide the minimum-bias (MB) trigger~\cite{Adler:2003zu} and 
are also used to determine the event centrality and vertex position 
along the beam axis ($z_{\rm BBC}$).  The MB trigger is formed if two or 
more BBC counters on each side detect the passage of charged 
particle(s).  The MB trigger efficiency in \cuau is 93\% of total 
inelastic collisions. The event centrality is defined by the total 
charge observed in the BBC. The mean number of participating nucleons 
(\Npart), binary collisions (\Ncoll) and the nuclear overlap function 
(\tab) in various centrality intervals are estimated with a 
Glauber-model Monte-Carlo simulation~\cite{Miller:2007ri} folded with 
the BBC response. The average values of \tab and \Npart for different 
centrality classes of \cuau collisions are listed in Table~\ref{tab0}.

The measurements are based on two data sets.  Up to moderate \pt 
($<8$\,\gevc) $6.9\times10^{9}$ MB events satisfying a vertex cut of 
$|z_{\rm BBC}|< 20$ cm are used. To improve statistics and extend the 
range to higher \pt an additional sample was collected with one of the 
EMCal hardware triggers (ERT-A). This 
trigger required the presence of at least one high-energy shower in the 
EMCal. After offline calibration it was found that the ERT-A trigger 
reached full efficiency for photons with energy above 4.5--5\,GeV 
depending on location in the calorimeter. The accumulated ERT-A data 
sample after the same $|z_{\rm BBC}|<20$\,cm vertex cut corresponds to 
$1.8\times10^{10}$ sampled MB events, which is a factor of three more 
than the MB sample. MB data is used to measure meson yields at $p_T<8$ 
GeV/$c$, and ERT-A data set is used at higher momenta.

Reconstruction of $\pi^0$ and $\eta$ mesons is performed via their decay 
modes $\pi^0\rightarrow\gamma\gamma$ and $\eta\rightarrow\gamma\gamma$. 
Photons are measured in the EMCal~\cite{Aphecetche:2003zr} located in the two central arms of the 
PHENIX detector, each covering 90 degrees in azimuth and $|\eta|<0.35$ 
in pseudorapidity. The EMCal comprises two technologically different 
subsystems: lead-scintillator sampling calorimeter (PbSc) and lead-glass 
\v{C}erenkov calorimeter (PbGl), which cover $3/8$ and $1/8$ of the full 
azimuth, respectively. The PbSc and PbGl subsystems have different 
linearity, energy resolution ($\delta E/E=2.1\%\oplus8.1\%/\sqrt{E}$ for 
PbSc and $0.8\%\oplus5.9\%/\sqrt{E}$ for PbGl) and segmentation 
($\delta\phi\times\delta\eta \approx 0.01\times0.01 $ for PbSc and $ 
0.008\times0.008 $ for PbGl).

\begin{figure}[t]
\includegraphics[width=1.0\linewidth]{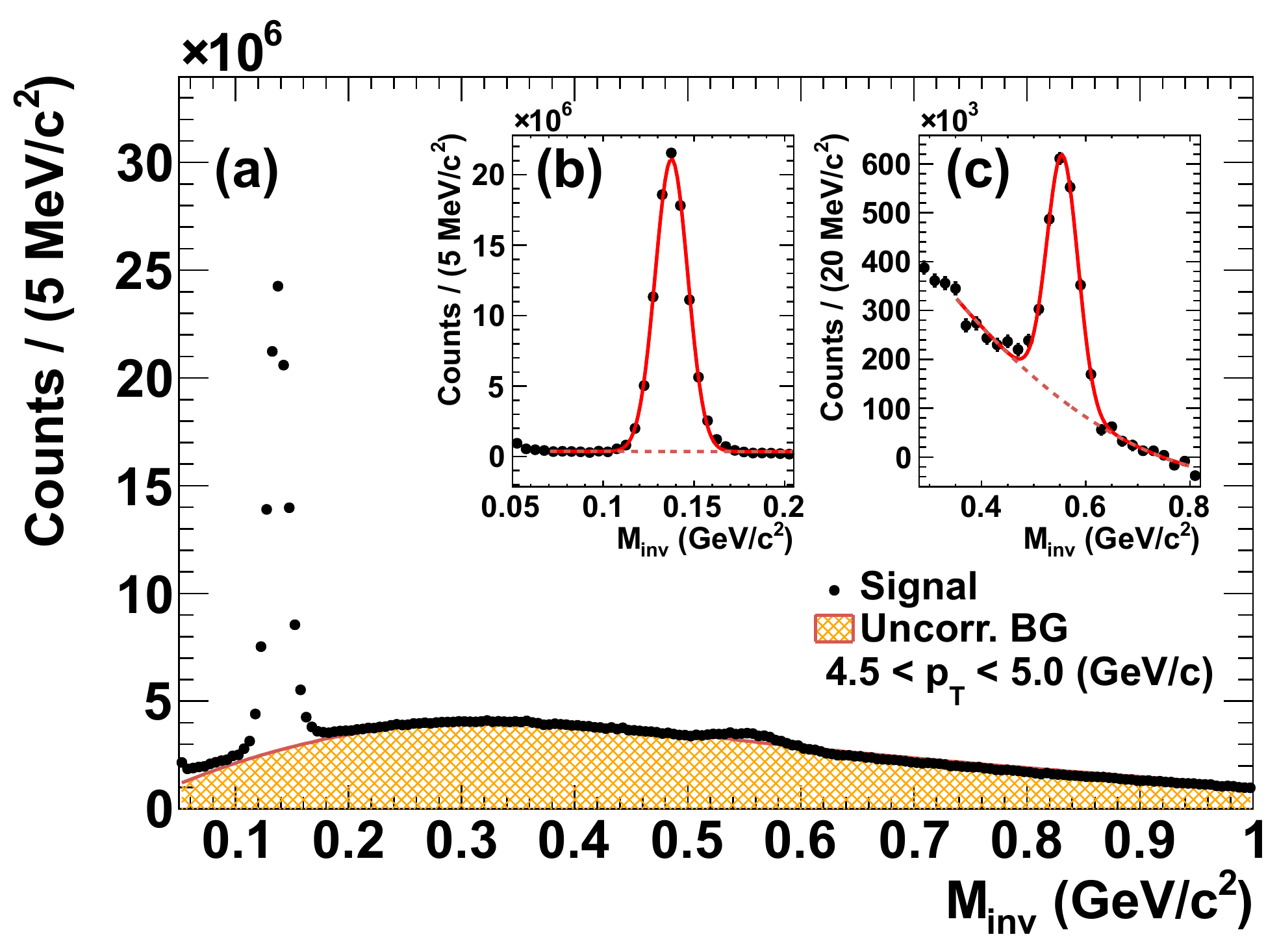}
\caption{Example of an invariant mass plot ($4.5<\pt<5.0$\,\gevc). (a) 
The foreground (photon pairs from the same event) is shown as points, 
the shaded area is the scaled mixed event background. Insert (b) shows 
the \piz peak area after mixed event subtraction; the Gaussian fit to 
the peak and the first order polynomial fit to the residual background 
are also shown.  Insert (c) shows the $\eta$ peak area with the Gaussian 
fit and second order polynomial for the residual background.
}
\label{fig:FIG0_InvMass_Example} 
\end{figure}

The raw yields of $\pi^0$ and $\eta$ mesons are determined from the 
$\gamma\gamma$ invariant mass ($m_{\rm inv}$) distribution, in bins of \pt 
and centrality. The analysis is carried out independently for the PbSc 
and PbGl subsystems. Photon candidates have to satisfy a shower shape 
cut~\cite{Aphecetche:2003zr} and are required to have energy 
($E_{\gamma}$) larger than 0.4\,GeV, which helps to further reduce the 
contribution from other particles, mostly minimum ionizing hadrons. Each 
$\gamma\gamma$ pair is required to satisfy an asymmetry cut 
$|\alpha|<0.8$, where 
$\alpha=|E_{\gamma1}-E_{\gamma2}|/(E_{\gamma1}+E_{\gamma2})$. The 
asymmetry cut helps to reduce the background from combinatorial 
$\gamma\gamma$ pairs in the $m_{\rm inv}$ distributions, improving the $S/B$ 
ratio. A typical invariant mass distribution is shown in 
Fig.~\ref{fig:FIG0_InvMass_Example}.

The $m_{\rm inv}$ distributions contain two peaks in the selected mass 
region, which correspond to decays of $\pi^0$ and $\eta$. At lower \pt 
the peaks sit on top of a large combinatorial background. The shape of 
the background is estimated by event mixing, i.e. from the $m_{\rm inv}$ 
distribution obtained by combining photons from different events that 
nevertheless have similar collision vertex and centrality. The mixed 
event distributions are normalized and subtracted from the real event 
distributions.  The mixed events are normalized outside of the meson 
peaks from $0.080<m_{\rm inv}<0.085$ and $0.3<m_{\rm inv}<0.4$ GeV/$c^2$ 
for the \piz and $0.7<m_{\rm inv}<0.8$\,\gevc for the $\eta$. The 
combinatorial background decreases rapidly with increasing \pt, 
therefore, mixed event subtraction is carried out only for $p_T$ below 
7--10\,\gevc depending on the collision centrality.  Above that the 
background under the peaks is estimated from the average counts in real 
events outside, but close to the peaks (sideband).

The resulting, combinatorial-subtracted invariant mass distributions are 
fit to a combination of a Gaussian to describe signal and a polynomial 
to describe the residual background. First- and second-order polynomials 
were used in $\pi^0$ and $\eta$ measurements, respectively. Meson raw 
yields were obtained as the difference between the integral of the bin 
content in the mass peak regions and the integral of the polynomial fits 
to the residual background in the same region. The mass peak regions 
were defined as $m_{\rm inv}=0.10$-$0.17$ and $0.48$-$0.62$ GeV/$c^2$ for 
$\pi^0$ and $\eta$, respectively.

Acceptance and reconstruction efficiency (efficiency hereafter) are 
estimated using a {\sc geant3}-based~\cite{Brun:1978fy} Monte-Carlo 
simulation of the PHENIX detector. The simulation was tuned to reproduce 
the observed mass peaks and widths of $\pi^0$ and $\eta$ in the real 
data. To account for the effect of underlying events the simulated 
mesons were embedded in real data in each centrality, then analyzed with 
the same methods as the real data. Final efficiencies also account for 
branching ratios of the analyzed decay modes and for the ERT-A trigger 
efficiency in the corresponding data sample.

Invariant yields of $\pi^0$ and $\eta$ are obtained as follows: 
\begin{equation}
   \frac{1}{N_{\rm event}}\frac{d^2N}{2\pi p_T dp_T dy} = 
\frac{N_{\rm raw}}{2\pi p_T N_{\rm event} \epsilon_{\rm rec} 
\Delta p_T \Delta y},
\end{equation}

where $N_{\rm raw}$ is the particle raw yield and $\epsilon_{\rm rec}$ 
is the efficiency (including acceptance and all other corrections), 
$N_{\rm event}$ is the number of analyzed events.

Systematic uncertainties are classified into three types. Type A 
represents uncertainties, that are entirely \pt-uncorrelated; these are 
added in quadrature to the statistical uncertainty. Type B uncertainties 
are \pt-correlated, but different from point to point, and all data 
points can move up or down by the same fraction of their Type B 
uncertainty. Type C represents uncertainties which move all points up or 
down by the same fraction. Typical values of the estimated systematic 
and total uncertainties are presented in Tables~\ref{tab1} and \ref{tab2}.

\begin{table*}[tbh]
\begin{minipage}{0.99\linewidth}
\caption{\label{tab1}
Systematic uncertainties for $\pi^0$ and $\eta$ yields at different \pt. 
Values are shown for PbSc(PbGl) subsystems. The types of uncertainties 
are described in the text. Values with a range indicate the variation of 
the uncertainty over the different centrality intervals.
}
\begin{ruledtabular} \begin{tabular}{cccccc}
Source & \multicolumn{2}{c}{$\pi^0\rightarrow\gamma\gamma$} & \multicolumn{2}{c}{$\eta\rightarrow\gamma\gamma$} & Type \\
 & $3.25$ GeV/$c$ & $11$ GeV/$c$ & $3.25$ GeV/$c$ & $11$ GeV/$c$ & \\
 \hline
Acceptance & 1.5$\%$(1.5$\%$) & 1.5$\%$(1.5$\%$) & 1.5$\%$(1.5$\%$) & 1.5$\%$(1.5$\%$) & B \\
$p_T$ weights & 1$\%$(1$\%$) & 1$\%$(1$\%$) & 1$\%$(1$\%$) & 1$\%$(1$\%$) & B \\
Energy scale & 5$\%$(5$\%$) & 9$\%$(9$\%$) & 5$\%$(5$\%$) & 9$\%$(9$\%$) & B \\
Energy resolution & 2$\%$(2$\%$) & 2$\%$(2$\%$) & 2$\%$(2$\%$) & 2$\%$(2$\%$) & B \\
ERT-A efficiency & $-$ & 1$\%$(1$\%$) & $-$ & 1.3$\%$(1.3$\%$) & B \\
Photon conversion & 5.2$\%$(5.2$\%$) & 5.2$\%$(5.2$\%$) & 5.2$\%$(5.2$\%$) & 5.2$\%$(5.2$\%$) & C \\
Cluster merging & $-$ & 3$\%$(2.5$\%$) & $-$ & $-$ & B \\
PID cuts & 4$\%$(4$\%$)$-$6$\%$(4$\%$) & 4$\%$(4$\%$)$-$6$\%$(4$\%$) & 5$\%$(5$\%$)$-$7$\%$(5$\%$) & 5$\%$(5$\%$)$-$7$\%$(5$\%$) & B \\
Raw yield extraction & 3$\%$(3$\%$) & 3$\%$(3$\%$)$-$4$\%$(4$\%$) & 8.13$\%$(8.13$\%$) & 5.3$\%$(5.3$\%$) & \\ 
Reconstruction efficiency & 0.8$\%$(1.5$\%$)$-$1.7$\%$(3$\%$) & 0.4$\%$(0.7$\%$)$-$0.7$\%$(1.5$\%$) & 3$\%$(5$\%$)$-$4$\%$(8$\%$) & 0.6$\%$(1.2$\%$)$-$1.03$\%$(1.9$\%$) & A \\
\end{tabular} \end{ruledtabular}
\end{minipage}
\begin{minipage}{0.99\linewidth}
\caption{\label{tab2}
Total uncertainties for $\pi^0$ and $\eta$ combined spectra, $R_{AB}$ and $\eta/\pi^0$ ratios at different \pt. The types of uncertainties are described in the text. Values with a range indicate the variation of the uncertainty over different centrality intervals.}
\begin{ruledtabular} \begin{tabular}{ccccc}
Type & $3.25$ GeV/$c$ & $11$ GeV/$c$ & $3.25$ GeV/$c$ & $11$ GeV/$c$ \\
\hline
 & \multicolumn{2}{c}{$\pi^0$ Combined Spectra} & \multicolumn{2}{c}{$\eta$ Combined Spectra} \\
Stat  & 0.08$\%-$0.2$\%$ & 0.9$\%-$4$\%$ & 3$\%-$5$\%$ & 4$\%-$11$\%$  \\
Type A & 0.9$\%-$1.8$\%$ & 0.4$\%-$0.9$\%$ & 2$\%-$4$\%$ & 0.6$\%-$0.9$\%$ \\
Type B & 6$\%-$7$\%$ & 10.1$\%-$10.3$\%$ & 8.7$\%-$9.3$\%$ & 10.7$\%-$11.1$\%$ \\
Type C & 5.2$\%$ & 5.2$\%$ & 5.2$\%$ & 5.2$\%$ \\
 \\
 & \multicolumn{2}{c}{$\pi^0$ $R_{AB}$} & \multicolumn{2}{c}{$\eta$ $R_{AB}$} \\
Type A + Stat. &  1.0$\%-$1.9$\%$ & 2.6$\%-$4.4$\%$ & 4$\%-$6$\%$ & 6$\%-$13$\%$ \\
Type B         &  10.5$\%-$10.7$\%$ & 14.3$\%-$14.4$\%$ & 13.7$\%-$14.1$\%$ & 14.4$\%-$14.7$\%$ \\
Type C         &  12$\%-$23$\%$ &  12$\%-$23$\%$ &  12$\%-$23$\%$ &  12$\%-$23$\%$ \\
 \\
 & \multicolumn{2}{c}{$\eta/\pi^0$} & \multicolumn{2}{c}{} \\
Type A + Stat. &  4$\%-$6$\%$ & 4$\%-$12$\%$ & & \\
Type B         &  10.9$\%-$11.4$\%$ & 14.7$\%-$15.1$\%$ & & \\
Type C         &  $-$ & $-$ & & \\
\end{tabular} \end{ruledtabular}
\end{minipage}
\end{table*}

One of the main sources of systematic uncertainties is the absolute 
energy calibration of the EMCal. The uncertainty 
on the absolute scale was estimated to be $1\%$. Due to the steeply 
falling (power-law) spectrum it corresponds to $\approx2$-$9\%$ uncertainty 
for the measured yields of $\pi^0$ and $\eta$ mesons, which gradually 
increases from low to high momentum. At high \pt the measured \piz 
yields are strongly affected by cluster merging when two photons from a 
$\pi^0$ decay have a small opening angle and produce partially or fully 
overlapping showers, which cannot be reconstructed as two individual 
clusters in the EMCal. Cluster merging results in significant loss of 
$\pi^0$ reconstruction efficiency at high $p_T$. Due to the different 
segmentation and Moliere-radius~\cite{Aphecetche:2003zr} the merging 
effect manifests itself differently in the PbSc and PbGl subsystems. In 
PbSc the merging starts at $p_T>12$ GeV/$c$, while in PbGl it starts 
only at $p_T>16$ GeV/$c$. Uncertainties on how well the simulations 
describe the merging effect result in corresponding uncertainties for 
the measured $\pi^0$ yields, increase with \pt, reaching $\approx$20$\%$ in 
PbSc and $\approx$9$\%$ in PbGl at 20 GeV/$c$.  Due to the four times 
heavier mass and larger $\gamma\gamma$ opening angle, the $\eta$ 
measurements will be influenced by cluster merging only at $p_T>50$ 
GeV/$c$, which is far beyond the $p_T$ range presented in this analysis. 
At low $p_T$ (below $\approx$5 GeV/$c$) the main uncertainty for $\pi^0$ 
and $\eta$ comes from the raw yield extraction due to relatively small 
$S/B$ ratios. This uncertainty is estimated as the maximum difference 
between raw yields obtained using different mass regions for mixed event 
background normalization, different fitting ranges, and different order 
polynomials for the residual background estimation. Some photons from 
$\pi^0$ and $\eta$ decays convert into $e^-e^+$ pairs when traversing 
through detector material. If this happens within the magnetic field, 
they are bent in opposite directions and can not be reconstructed as a 
single photon-like cluster in the EMCal. As a result, $\approx25\%$ of 
$\pi^0$ and $\eta$ mesons are lost. This effect is included in the 
efficiency calculation.  The uncertainty on how accurately it is 
reproduced in the simulation is estimated to be $5.2\%$, and it is 
Type-C, because in the relevant energy range the conversion probability is 
almost constant.

Systematic uncertainties for $\eta/\pi^0$ ratios are included as a 
quadratic sum of the type-B uncertainties from $\pi^0$ and $\eta$ 
yields. Because type-C uncertainties of the $\pi^0$ and $\eta$ yields are 
100\% correlated between these particle measurements for all \pt, this 
uncertainty cancels in the ratios.  The $p_T$-correlated systematic 
uncertainties for $R_{AB}$ include both uncertainties from Cu$+$Au and 
$p$+$p$ measurements~\cite{Adare:2007dg}.

\begin{figure*}[htbp]
  \includegraphics[width=0.98\linewidth]{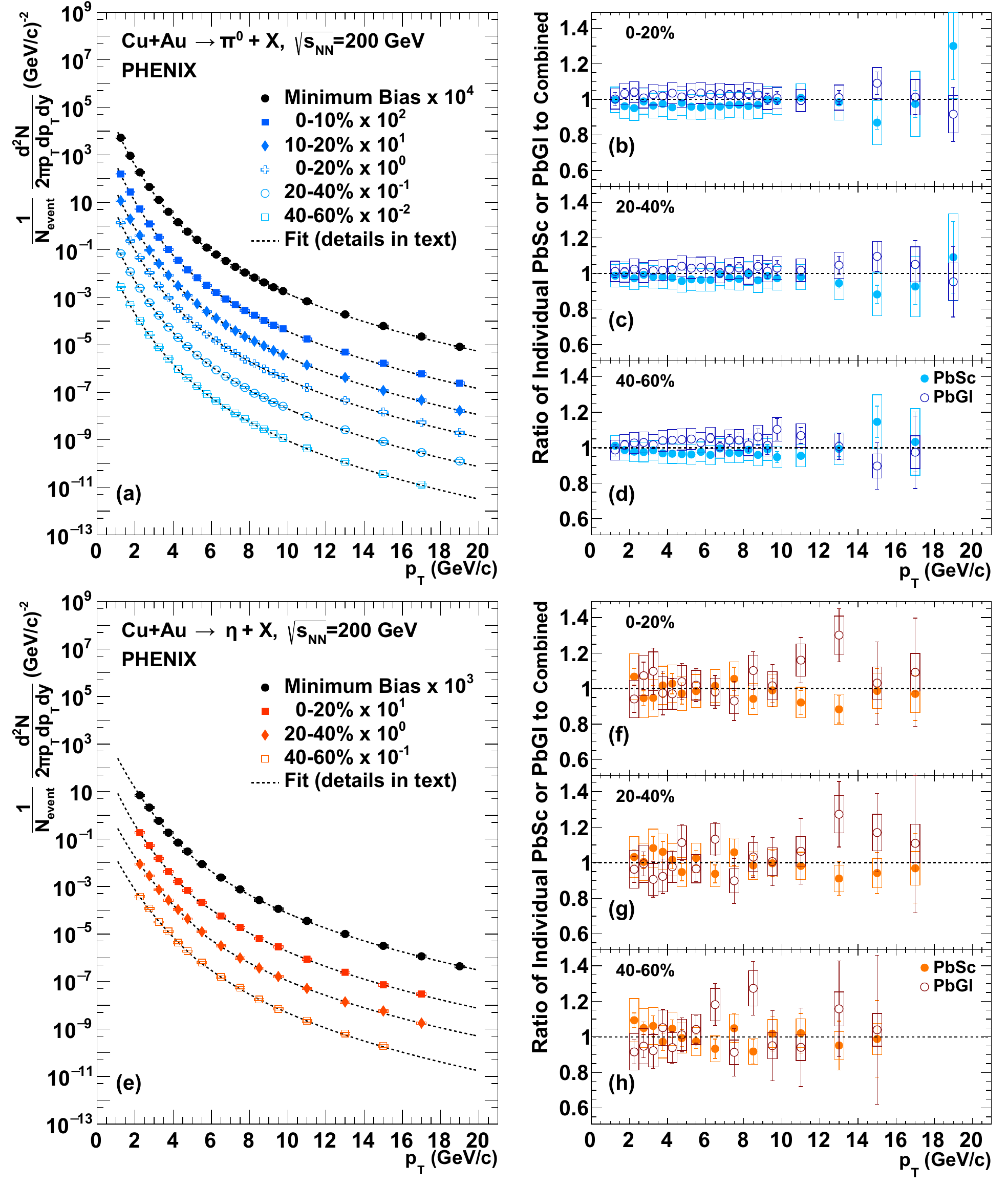}
\caption{Left: $\pi^0$ (a) and $\eta$ (e) invariant $p_T$-spectra 
measured in different centrality intervals of $\cuau$ collisions at 
$\sqrt{s_{_{NN}}}$=200 GeV. The dashed curves are a fit with two 
Hagedorn-type functions with an asymptotic power-law ($\pt^{-n}$) 
behavior. Right: ratios of $\pi^0$ (b-d) and $\eta$ (f-h) yields 
measured in PbSc or PbGl subsystem to the averaged ones. Error bars 
represent a quadratic sum of statistical and type-A systematic 
uncertainties. Error boxes in the right panel correspond to the 
quadratic sum of systematic uncertainties, which are uncorrelated 
between PbSc and PbGl.
}
\label{fig:FIG1_CuAu_CY}
\end{figure*}

Invariant yields are obtained separately for PbSc and PbGl subsystems. 
The results are then averaged with weights defined by the quadratic sum 
of statistical and those systematic uncertainties that are uncorrelated 
between the two subsystems. The ratios of the yields obtained in PbSc 
and PbGl to the averaged ones are presented in panels (b)-(d) and 
(f)-(h) of Fig.~\ref{fig:FIG1_CuAu_CY}. Only uncorrelated systematic 
uncertainties are shown in the ratios. Yields obtained in the different 
subsystems are consistent within statistical and uncorrelated systematic 
uncertainties.  Typical systematic uncertainties for the combined 
spectra, \piz and $\eta$ $R_{AB}$ and $\eta$/\piz ratio are listed in 
Table~\ref{tab2}.

To facilitate comparison between different experiments and data sets, the 
data points of the meson spectra are plotted at the center of each given 
$p_T$ interval, which, due to the falling spectrum, does not represent 
the true physical value of the yield at that \pt~\cite{LAFFERTY1995541}. 
A bin-shift correction is applied that adjusts the meson yields to their 
value at the bin center.

\begin{figure}[bp]
\includegraphics[width=1.0\linewidth]{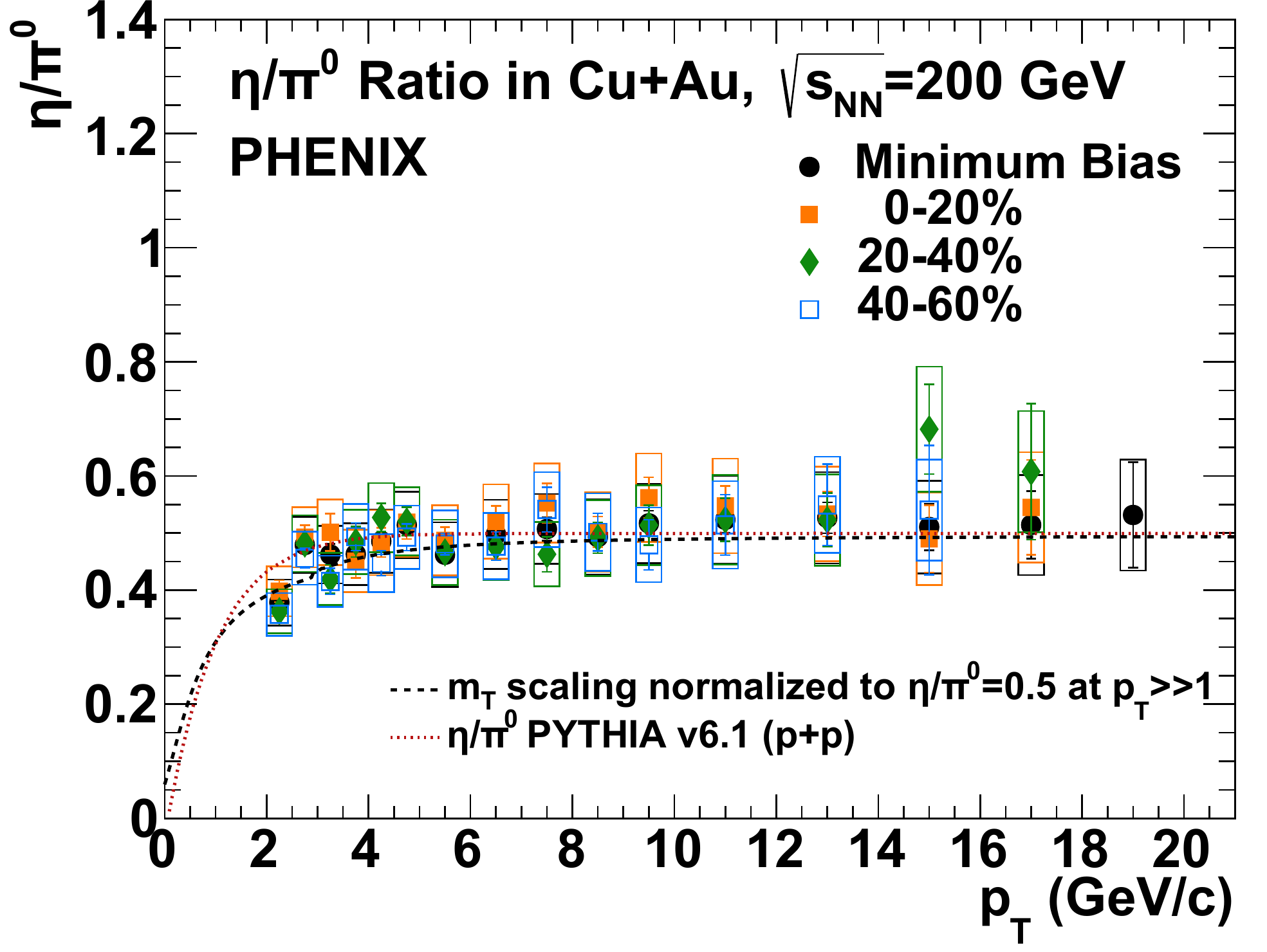}
\caption{ Ratios of $\eta$ and $\pi^0$ yields measured as a function of 
$p_T$ in different centrality intervals of Cu$+$Au collisions at 
$\sqrt{s_{_{NN}}}$=200 GeV. Dashed curve shows the $m_T$ scaling curve 
normalized to 0.5 at high momentum (see text). Error bars represent a 
quadratic sum of statistical and type-A systematic uncertainties for 
$\pi^0$ and $\eta$ yields. Error boxes represent a quadratic sum of 
type-B systematic uncertainties for $\pi^0$ and $\eta$ yields.
}
\label{fig:FIG2_CuAu_eta_pi0_rat}
\end{figure}

\begin{figure*}[th]
\begin{minipage}[tbh]{0.6\linewidth}
\includegraphics[width=0.99\linewidth]{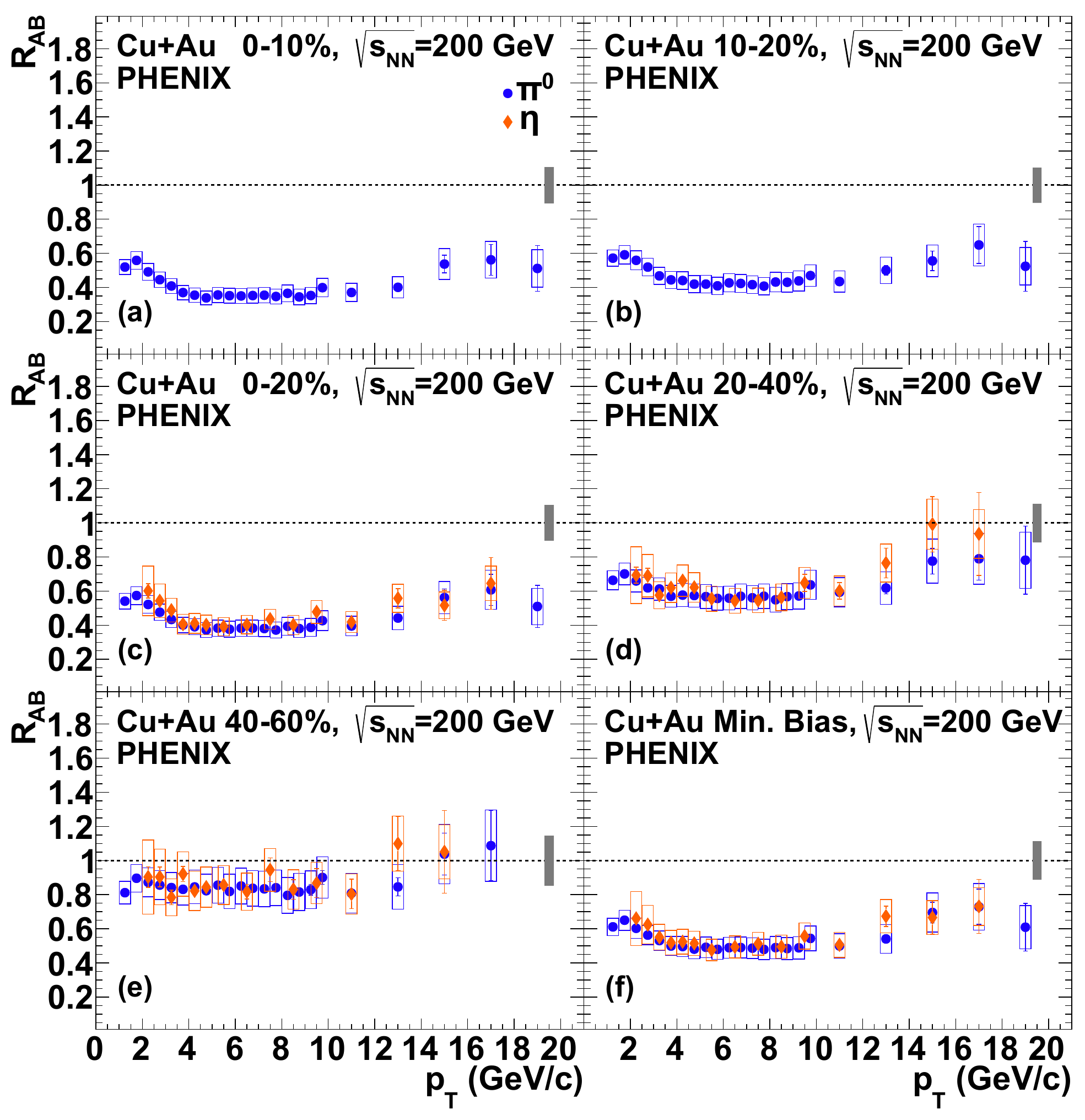}
\end{minipage}
\begin{minipage}[tbh]{0.3\linewidth}
\hspace{1.0cm}
\caption{ $R_{AB}$ of $\pi^0$ and $\eta$ mesons measured as a function 
of $p_T$ in different centrality intervals of Cu$+$Au collisions at 
$\sqrt{s_{_{NN}}}$=200 GeV. Error bars represent a quadratic sum of 
statistical and type-A systematic uncertainties from Cu$+$Au and $p$+$p$ 
measurements, respectively. Error boxes represent type-B systematic 
uncertainties from Cu$+$Au and $p$+$p$ measurements. Solid and open boxes 
at unity represent type-C systematic uncertainties from Cu$+$Au (including 
uncertainties from the $T_{AB}$ values) and $p$+$p$, respectively. The 
reference $p$+$p$ measurements are published in~\cite{Adare:2007dg} for 
$\pi^0$ and in~\cite{Adare:2010cy,Adare:2010dc} for $\eta$ (see details 
in the text).
}   
\label{fig:FIG3_CuAu_pi0_eta_RAA}
\end{minipage}
\end{figure*}

\begin{figure*}[tbh]
\begin{minipage}[tbh]{0.46\linewidth}
\includegraphics[width=0.99\linewidth]{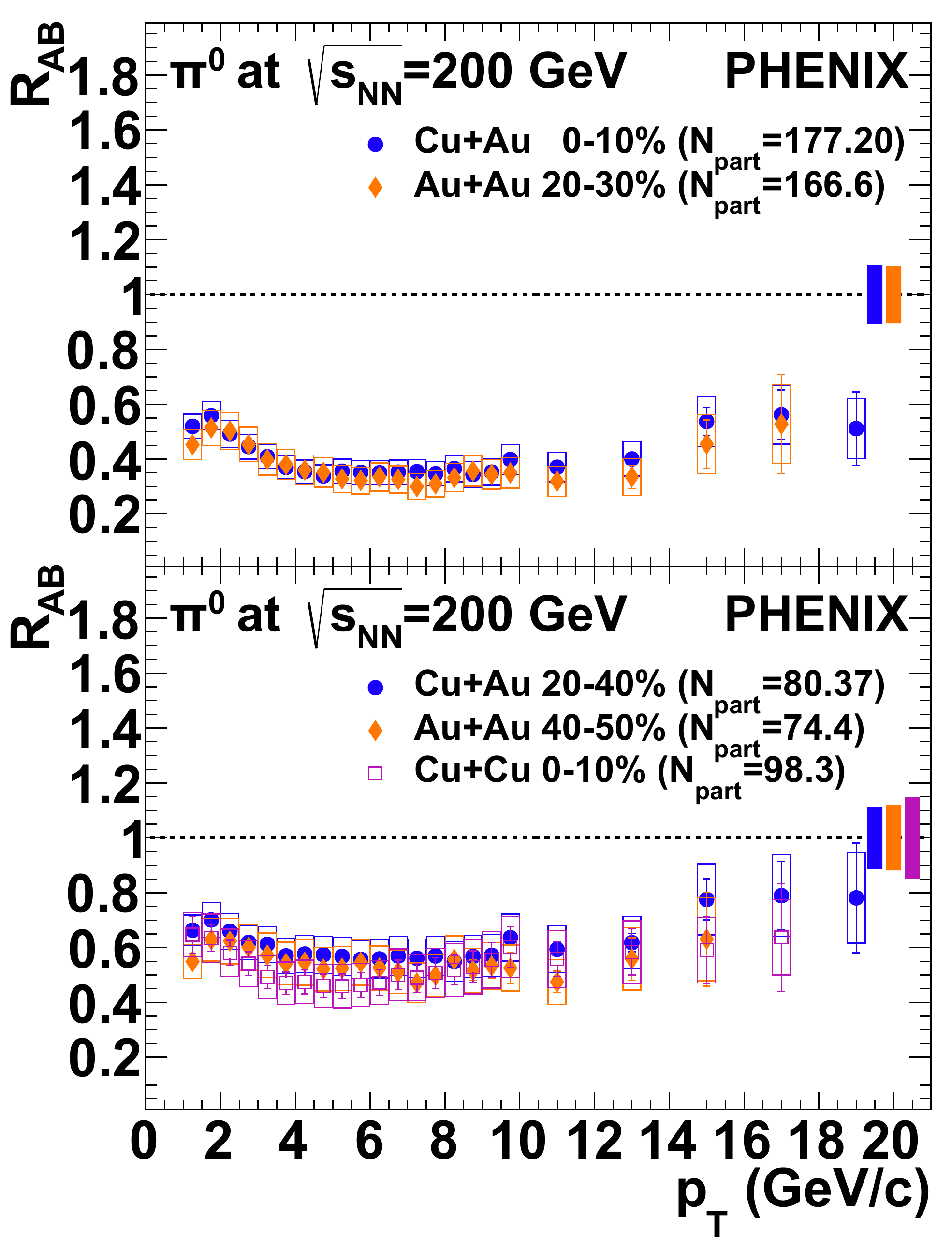}
\caption{Comparison of $\pi^0$ $R_{AB}$ measured in $\cuau$, $\auau$ 
and $\cucu$ collisions at $\sqrt{s_{_{NN}}}=200$ GeV and comparable \Npart. 
Error bars represent a quadratic sum of statistical and type-A 
systematic uncertainties from Cu$+$Au and $p$$+$$p$ measurements. Open boxes 
are Type-B systematic uncertainties for Cu$+$Au and \pp collisions. 
The three boxes at unity are type-C systematic 
uncertainties from $p$$+$$p$ and heavy-ion collisions. The boxes from 
left to right correspond to Cu$+$Au, Au$+$Au and Cu$+$Cu measurements, 
respectively.
}
\label{fig:fig5}
\end{minipage}
\hspace*{0.2cm}
\begin{minipage}[tbh]{0.46\linewidth}
\vspace{-1.75cm}
\includegraphics[width=0.99\linewidth]{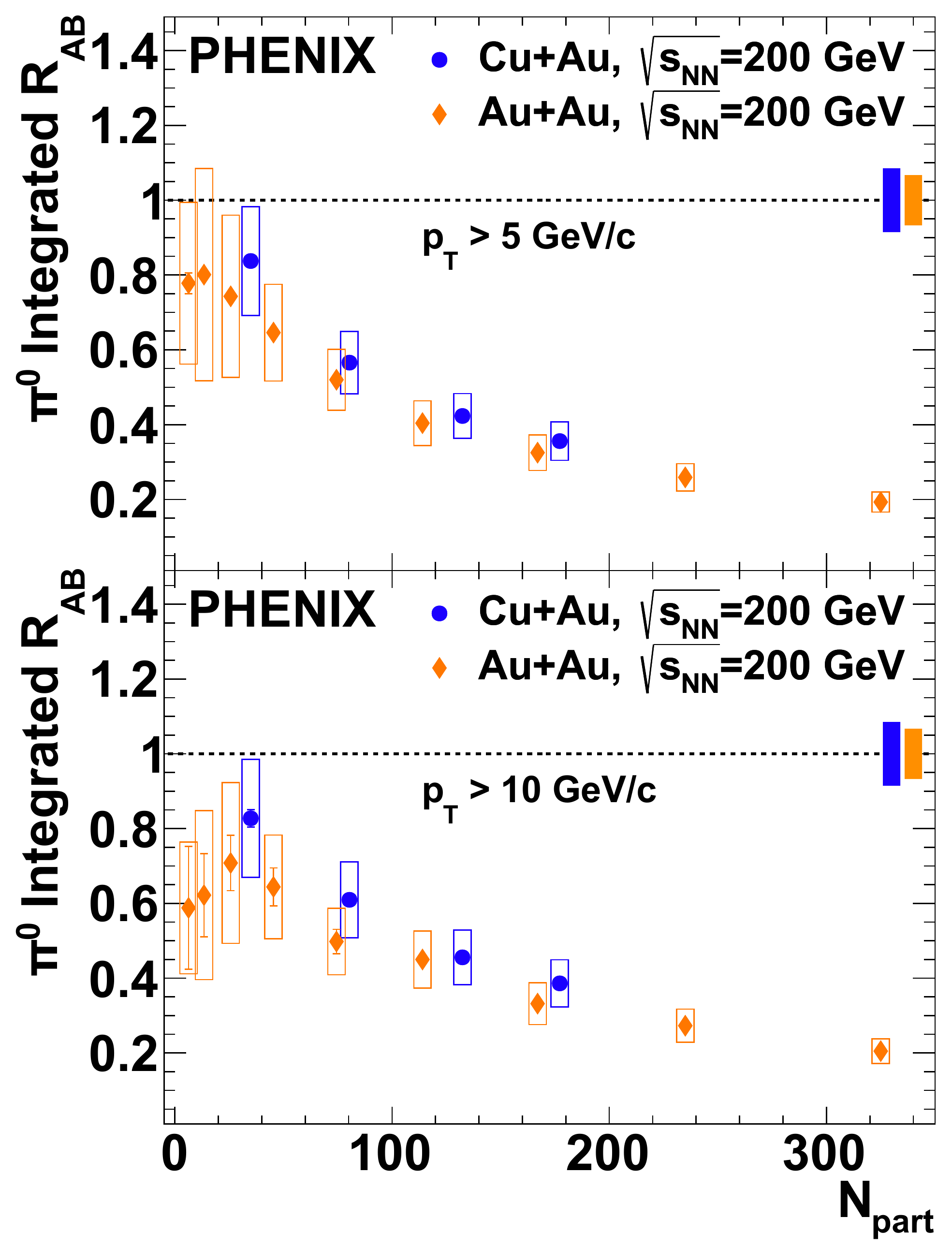}
\caption{ Comparison of integrated $R_{AB}$ for \piz measured as a 
function of $N_{\rm part}$ in $\cuau$ and $\auau$ collisions at 
$\sqrt{s_{_{NN}}}=200$ GeV. Uncertainties are the same as in the 
Fig.~\ref{fig:fig5}.  The lower limit of integration is \pt=5\,\gevc on 
panel (a), \pt=10\,\gevc on panel (b).
}
\label{fig:fig6}
\end{minipage}
\end{figure*}

\section{Results and discussion}

Invariant yields in the \pt range 1(2)-20\,\gevc for $\pi^0$ ($\eta$) 
mesons measured in different centrality intervals and MB 
collisions are shown in panels (a) and (e) of 
Fig.~\ref{fig:FIG1_CuAu_CY}, respectively. At low \pt the measurement is 
limited by the rapidly decreasing $S/B$ ratio, and at high \pt by the 
available statistics.

Spectra of $\pi^0$ and $\eta$ mesons can be fitted with a sum of 
Hagedorn and power-law functions:
\begin{equation}
   f(p_T)=T(p_T)\frac{A}{(1+p_T/p_0)^m}+(1-T(p_T))\frac{B}{p^n_T},
\end{equation}

\noindent where $T(p_T)=1/(1+\exp((p_T-t)/w))$, $A$, $p_0$, $m$, $B$, 
$n$, $t$ and $w$ are free parameters. The parameter $t$ governs at what 
\pt the second, pure power-law term becomes dominant; $t$ varies between 
4--6\,\gevc, depending on centrality. The parameter $w$ varies between 
0.05--0.15\, \gevc and governs the width of transition interval, where 
the first term loses its dominance and the second term becomes dominant. 
At high transverse momenta $f(p_T) \propto \pt^{-n}$.  For \piz in MB 
collisions $n=8.06 \pm 0.01_{\rm stat} \pm 0.06_{\rm sys}$, for the most 
central 0\%--10\% collisions 
$n=8.02 \pm 0.02_{\rm stat} \pm 0.07_{\rm sys}$, and increases slowly to 
$n=8.07 \pm 0.02_{\rm stat} \pm 0.06_{\rm sys}$ up to 40\% centrality. 
These numbers are consistent within uncertainties to the values obtained 
in pure power-law fits at high \pt ($>$8\,\gevc) in 200\,GeV \auau 
collisions with similar \Npart~\cite{Adare:2008qa,Adare:2012wg}.

The $\eta/\pi^0$ ratios ($R_{\eta/\pi^0}$) as a function of $p_T$ for 
different Cu$+$Au centrality intervals are presented in the 
Fig.~\ref{fig:FIG2_CuAu_eta_pi0_rat}. Within uncertainties the measured 
$R_{\eta/\pi^0}$ are centrality independent in the whole $p_T$ range of 
measurements.  A constant fit to the MB data in the 
4$<\pt<$20\,\gevc region results in $\eta/\piz = 0.50 \pm 0.01_{\rm stat} 
\pm 0.02_{\rm sys}$, and the various centrality bins are consistent with 
this value.  The dashed curve in Fig.~\ref{fig:FIG2_CuAu_eta_pi0_rat} 
shows this asymptotically constant fit modified according to 
$m_T$-scaling. Similar results were obtained in hadron-hadron, 
hadron-nucleus, and nucleus-nucleus collisions as well as in 
$e^{+}e^{-}$ collisions in a wide range of collision energies 
$\sqrt{s_{_{NN}}}=3$--1800 GeV~\cite{Busser:1974yj,Adler:2006bv}. This 
suggests that QGP medium produced in Cu$+$Au collisions either does not 
affect the jet fragmentation into light mesons or it affects the \piz 
and $\eta$ the same way.

Nuclear modification factors of $\pi^0$ and $\eta$ mesons as functions 
of $p_T$ are shown in Fig.~\ref{fig:FIG3_CuAu_pi0_eta_RAA} for different 
Cu$+$Au centrality intervals. The reference $\pi^0$ meson production 
cross section in $p$+$p$ collisions was obtained from the 2005 PHENIX 
$p$+$p$ measurement~\cite{Adare:2007dg}. For $\eta$ meson $R_{AB}$ 
estimation, the 2006 PHENIX $p$+$p$ measurements were 
used~\cite{Adare:2010cy}. The $R_{AB}$-s of $\pi^0$ and $\eta$ mesons 
are consistent within uncertainties in the whole $p_T$ range for every 
analyzed centrality interval of Cu$+$Au collisions. At $p_T>5$ GeV/$c$ 
$R_{AB}$ is $\approx 0.4 - 0.5$ in most central collisions. A weak $p_T$ 
dependence of the measured $R_{AB}$ values can be observed. The 
suppression of $\pi^0$ and $\eta$ decreases as one moves to more 
peripheral collisions.

Fig.~\ref{fig:fig5} compares $R_{AB}$ of $\pi^0$ mesons measured as a 
function of $p_T$ in $\cuau$, $\auau$~\cite{Adare:2008qa} and 
$\cucu$~\cite{Adare:2008ad} collisions at $\sqrt{s_{_{NN}}}=200$ GeV and 
similar $N_{\rm part}$. In central and semi-central $\cuau$ collisions 
$\pi^0$ $R_{AB}$ are consistent with those measured in $\auau$ and 
$\cucu$, if applicable, which suggests that $\pi^0$ suppression mostly 
depends on the energy density and size of the produced medium.  Because in 
the most central collisions the Cu ion is fully submerged in Au, without 
any "corona"~\cite{Pantuev:2005jt}, but the suppression is the same as 
in \auau at comparable \Npart, the corona-effect is either nonexistent 
or very small.

In Fig.~\ref{fig:fig6}, the $\pi^0$ and $\eta$ integrated $R_{AB}$'s are 
shown as a function of $N_{\rm part}$ and compared to $\auau$. The 
integration is carried out in two different \pt ranges ($\pt>5$\,\gevc 
and $\pt>10$\,\gevc). The results obtained for the two different 
collision systems are consistent within uncertainties.

\section{Summary}

In summary, PHENIX has measured $\pi^0$ and $\eta$ invariant 
$p_T$-spectra and nuclear modification factors in asymmetric collisions 
of heavy ions, Cu$+$Au at $\sqrt{s_{_{NN}}}=200$ GeV in a wide $p_T$ range 
($1(2)<p_T<20$ GeV/$c$) and for several centrality intervals. In the 
more central collisions the spectra are similar to those observed in 
\auau. The asymptotic (high \pt) value of $\eta/\piz$ is $0.50 \pm 
0.01_{\rm stat} \pm 0.02_{\rm sys}$, constant, independent of collision 
centrality, and consistent with the previously measured values in 
hadron-hadron, hadron-nucleus, nucleus-nucleus as well as $e^{+}e^{-}$ 
collisions at $\sqrt{s_{_{NN}}}=$3--1800 GeV, suggesting that either the 
fragmentation of jets into \piz and $\eta$ is unchanged irrespective of 
the absence or presence of the medium, or it changes the same way, 
despite the different flavor content. The values of $R_{AB}$ for \piz 
and $\eta$ are consistent within uncertainties in all analyzed 
centrality intervals of $\cuau$ collisions. The suppression pattern of 
$\pi^0$ in $\cuau$ collisions is consistent with $\auau$ and $\cucu$ 
collisions at the same interaction energy and similar values of 
$N_{\rm part}$.

\begin{acknowledgments}


We thank the staff of the Collider-Accelerator and Physics
Departments at Brookhaven National Laboratory and the staff of
the other PHENIX participating institutions for their vital
contributions.  We acknowledge support from the 
Office of Nuclear Physics in the
Office of Science of the Department of Energy,
the National Science Foundation, 
Abilene Christian University Research Council, 
Research Foundation of SUNY, and
Dean of the College of Arts and Sciences, Vanderbilt University 
(U.S.A),
Ministry of Education, Culture, Sports, Science, and Technology
and the Japan Society for the Promotion of Science (Japan),
Conselho Nacional de Desenvolvimento Cient\'{\i}fico e
Tecnol{\'o}gico and Funda\c c{\~a}o de Amparo {\`a} Pesquisa do
Estado de S{\~a}o Paulo (Brazil),
Natural Science Foundation of China (People's Republic of China),
Croatian Science Foundation and
Ministry of Science and Education (Croatia),
Ministry of Education, Youth and Sports (Czech Republic),
Centre National de la Recherche Scientifique, Commissariat
{\`a} l'{\'E}nergie Atomique, and Institut National de Physique
Nucl{\'e}aire et de Physique des Particules (France),
Bundesministerium f\"ur Bildung und Forschung, Deutscher Akademischer 
Austausch Dienst, and Alexander von Humboldt Stiftung (Germany),
J. Bolyai Research Scholarship, EFOP, the New National Excellence
Program ({\'U}NKP), NKFIH, and OTKA (Hungary),
Department of Atomic Energy and Department of Science and Technology 
(India),
Israel Science Foundation (Israel), 
Basic Science Research Program through NRF of the Ministry of 
Education (Korea),
Physics Department, Lahore University of Management Sciences (Pakistan),
Ministry of Education and Science, Russian Academy of Sciences,
Federal Agency of Atomic Energy (Russia),
VR and Wallenberg Foundation (Sweden), 
the U.S. Civilian Research and Development Foundation for the
Independent States of the Former Soviet Union, 
the Hungarian American Enterprise Scholarship Fund,
the US-Hungarian Fulbright Foundation,
and the US-Israel Binational Science Foundation.

\end{acknowledgments}


%
 
\end{document}